\documentclass[pra,twocolumn,showpacs]{revtex4}
\usepackage{amssymb,amsmath,dsfont,graphicx}

\newcommand{\Ham}{\mathcal{\hat H}}

\newcommand{\be}{\begin{eqnarray}} 
\newcommand{\ee}{\end{eqnarray}}

\begin{document} 
%%%%%%%%%%%%%%%%%%%%%%%%%%%%%%%%%%%%%%%%%%%%%%%%%%%%%%%%%%%%%%%%%%%%%%%%%%%%%% 
%%%%%%%%%%%%%%%%%%%%%%%%%%%%%%%%%%%%%%%%%%%%%%%%%%%%%%%%%%%%%%%%%%%%%%%%%%%%%% 
\title{Stochastic Simulation of a finite-temperature one-dimensional Bose-Gas: 
from Bogoliubov to Tonks-Girardeau regime} 
\author{B. Schmidt, L.I. Plimak, and M. Fleischhauer} 
\affiliation{Fachbereich Physik, Technische Universit{\"a}t Kaiserslautern,  
D-67663 Kaiserslautern, Germany} 
%%%%%%%%%%%%%%%%%%%%%%%%%%%%%%%%%%%%%%%%%%%%%%%%%%%%%%%%%%%%%%%%%%%%%%%%%%%%%% 
\begin{abstract} 
We present an ab initio stochastic method for calculating thermal properties  
of a trapped, 1D Bose-gas covering the whole range from weak to strong  
interactions. Discretization of the problem results in a 
Bose-Hubbard-like Hamiltonian, whose imaginary time evolution  
is made computationally accessible by stochastic factorization of 
the kinetic energy. To achieve convergence for low enough temperatures such  
that quantum fluctuations are essential, the stochastic 
factorization is generalized to blocks, and ideas from  
density-matrix renormalization are employed.   
We compare our numerical results for density and first-order correlations 
with analytic predictions.  
% Our results are consistent with the  
% transition from thermal, i.e.  
% exponential, to Luttinger-liquid like, i.e.  
% power-law decay of correlations. 
\end{abstract} 
%%%%%%%%%%%%%%%%%%%%%%%%%%%%%%%%%%%%%%%%%%%%%%%%%%%%%%%%%%%%%%%%%%%%%%%%%%%%% 
%%%%%%%%%%%%%%%%%%%%%%%%%%%%%%%%%%%%%%%%%%%%%%%%%%%%%%%%%%%%%%%%%%%%%%%%%%%%%% 
\maketitle 
%\section{Introduction} 
%\begin{figure} 
%\scalebox{1}{\includegraphics{Pic/mu6g25.eps}} 
%\end{figure} 
 
For a long time the homogeneous 1D gas was a rather 
academic toy model, mainly of interest for mathematical physics  
since it is analytically tractable using the Bethe Ansatz  
\cite{lieb1963,yang1968,bogoliubov-book}. Recent  
successes in trapping and cooling bosonic atoms  
have lead  however to a number of 
experiments culminating in the observations of 
Tonks-gas properties in lattices \cite{paredes2004} and harmonic traps  
\cite{Kinoshita2004}. These developments have also lead to a  
rapidly growing theoretical interest in  trapped  
1D gases, which 
can be described  
by the grand canonical Hamiltonian  
%$\hat {\mathcal K} 
%=\Ham-\mu {\mathcal N}$  
\cite{olshanii1998}: 
\begin{multline} 
\label{Hamiltonian} 
\hat{\mathcal K}=\int\!\! 
dx\ \hat\psi^\dagger(x)\left[ 
 -\frac{\hbar^2}{2m}\partial_x^2 + V(x) - \mu  
\right]\hat\psi(x)  
\\ + \frac{g_\text{1D}}{2}\int\!\! dx\!\!\ 
\hat\psi^{\dagger}(x)\hat\psi^{\dagger}(x)\hat\psi(x)\hat\psi(x) 
. 
\end{multline} 
Here $m$ is the mass of the bosons, $g_{\rm 1D}$---% 
the 1D interaction constant, $V(x)$---the (harmonic) confining potential,  
and $\mu$---the chemical 
potential.  
 
Physics of degenerate quantum gases in one dimension is dominated 
by quantum fluctuations and thus differs considerably from that in higher 
dimensions. Homogeneous bosons  
do not form a condensate in 1D 
\cite{Hohenberg1967}. At any $T>0$ first-order 
correlations decay exponentially with distance; a power-law decay 
occurs only at $T=0$ \cite{efetov1975,Haldane1981}.  
The most peculiar feature of 1D interacting bosons is that they become 
less ideal with decreasing density, $n$: the bosons start  
avoiding each other as if they were noninteracting fermions.  
This regime, called the  
Tonks-Girardeau gas \cite{Tonks1936,Girardeau1960},  
is reached if the Tonks 
parameter, 
$ \gamma={m g_{\rm 1D}}/{\hbar^2 n}$, is large compared to unity. 
 
Experimentally, a quasi-1D regime  
can be realized in highly anisotropic traps, where the radial motion is  
confined to zero point oscillations and thus effectively eliminated. 
In the presence of a 1D trapping potential, exact solutions  
are no longer available.  
Local properties can be derived in a local density approximation from the 
Bethe-Ansatz solutions \cite{Kheruntsyan2003}. 
The finite size of the trapped system results in an energy gap, 
which allows for the formation 
of a quasi-condensate in the weakly interacting regime $\gamma\ll 1$. 
In this limit the quasi-condensate density 
can be obtained from the Gross-Pitaevskii equation, and  
correlations can be calculated  
within Bogoliubov and Thomas-Fermi  
approximations \cite{petrov2000,luxat2003,olshanii2003}. 
 
Long-distance correlations in the homogeneous gas can be described 
for all values of $\gamma$ using  
an effective harmonic-fluid  
(Luttinger liquid---LL) model  
\cite{efetov1975,Haldane1981,Cazalilla-JPhysB-2004}, which  
is analytically tractable.  
At $T=0$ it yields  
an algebraic decay of phase correlations $\sim 1/x^{1/2 K}$,  
where 
$K\ge 1$ is the so-called Luttinger parameter. In the case of a trap the  
1D gas can also be related to a LL model 
\cite{Monien1998} and solved within a local density approximation 
\cite{Gangardt2003}.  
In a trap, the power law   
sets in at temperatures of the order of the trap frequency,  
$k_\text{B}T\approx\hbar\omega$. 
 
We here present a stochastic method to numerically calculate  
the quantum properties of a trapped finite-temperature 1D Bose gas. 
We are particularly interested in the transition from the Gross-Pitaevskii 
to the Tonks-Girardeau regime as manifested in the behavior of the 
density and first order correlations. So as to achieve convergence 
for temperatures as low as the characteristic phase temperature $T_{\rm ph}$ 
introduced in \cite{petrov2000}, we employ an extension of 
the stochastic Gutzwiller approach for the 
grand canonical density operator \cite{Plimak-DPG-2003,carusotto-NJPh-2003}. 
Like the stochastic Gutzwiller Ansatz, our method 
is based on a stochastic factorization of the kinetic energy of a 
discretized Hamiltonian, which is however  
applied to {\em blocks\/} of sites.  
Furthermore, the blocks are  
treated by methods inspired by density matrix 
renormalization group (DMRG) techniques \cite{Schollwoeck2004}.   
In the strong 
interaction regime the kinetic energy is effectively a perturbation,  
which makes our approach  
more suitable than, e.g., the  stochastic Hartree approach  
\cite{Carusotto-PRL-2003}, (which has been 
very successful in describing the occupation number  
statistics of a weakly interacting BEC)  
or  phase-space techniques  
of quantum optics in  
imaginary time \cite{drummond1980,plimak1998}.  
Recent extensions of  
the latter involving stochastic gauges  
\cite{Drummond2004} also allow access to the regime 
of strong interactions, yet  
seem to be limited to higher temperatures.

We start from deriving  
a consistently discretized version of (\ref{Hamiltonian}). 
Let equidistant  
grids in position- and quasi-momentum spaces be introduced,  
with grid constants  
$\Delta x$ and $\Delta p$, respectively:    
$ \Delta x\Delta p={2\pi\hbar}/{M}$, $M$ being the number of grid points. 
This is equivalent to putting the system into a box of size $L=M\Delta x$  
with periodic boundary conditions and restricting the quasi-momentum to  
an interval of length $P=M\Delta p$. 
The Bose field with discretized modes 
$\hat a_j$ corresponding to wave numbers $k_j=p_j/\hbar$ 
is related  
to local bosonic operators 
via the discrete Fourier transformation:  
($j,l = 0,\cdots,M-1$) 
\begin{eqnarray} 
\hat b_j=\frac{1}{\sqrt{M}}\sum_{l=0}^{M-1}\hat a_{l}e^{ik_{l}x_j}, 
\quad [\hat b_j,\hat b_l^\dagger]=\delta_{jl}. 
\label{localop} 
\end{eqnarray} 
In terms of the discretised fields Hamiltonian  
(\ref{Hamiltonian}) reads 
\begin{gather} 
\begin{gathered} 
\hat{\mathcal K}=\Ham_\text{kin}+\Ham_0, \ \ \  
\Ham_\text{kin} = \frac{\hbar^2}{2m}\sum_{ij}C_{ij}\hat b^\dagger_i \hat b_{j},  
\\  
\Ham_0 = \sum_j\hat 
b_j^\dagger\hat b_j (V_j-\mu)+\frac{g_\text{1D}}{2\Delta x} 
\sum_j\hat b_j^{\dagger 2}\hat b_j^2. 
\end{gathered} 
\label{discrH} 
\end{gather} 
For symmetric $p_j$ and sufficiently small $\Delta x$ one finds 
\begin{equation} 
\label{nnhop} 
C_{ij}=\frac{1}{\Delta x^2}(2\delta_{ij}-\delta_{ij+1}-\delta_{ij-1}) 
\end{equation} 
which only contains nearest-neighbor hopping. 
Importantly, the kinetic energy matrix scales like $\Delta x^{-2}$,  
whereas the interaction---like $\Delta x^{-1}$.  
 
The discretized grand canonical  
Hamiltonian is equivalent 
to a Bose-Hubbard model (BHM) with effective hopping  
$J=\hbar^2/2 m \Delta x^2$, effective on-site interaction  
$U=g_{\rm 1D}/\Delta x$, and effective 
chemical potential $\mu_{\rm BH}=\mu-2 J$.  
The scaled hopping can be expressed in terms of the Tonks parameter 
at the trap center,  
${J}/{U} = {2}/{\gamma n(0) \Delta x}$. 
%%%%%%%%%%%%%%%%%%%%%%%%%%%%%%%%%%%%%%%%%%%%%%%%%%%%%%%%%
%
%  new text in revised version 04.02.2005
%
For fixed parameters, the discretized model is identical to the
continuous one for $\Delta x \to 0$. In this limit always the weak coupling
case $J/U\to 0$ of the corresponding Bose-Hubbard model is recovered, 
which prohibits an insulating phase. Thus the 
1D gas corresponds to a superfluid phase of the BHM close to the 
line $\mu_{\rm BH}=-2 J$. 
However, for finite values of $\Delta x$, which should be 
less than $n(0)^{-1}$ and small enough in order to stay outside the $n=1$
insulator lobe, a strong interaction limit $J/U\ll 1$ is still 
possible for large values of
$\gamma$. 
%
%%%%%%%%%%%%%%%%%%%%%%%%%%%%%%%%%%%%%%%%%%%%%%%%%%%%%%%%%%

Our goal is to numerically evaluate the grand canonical  
statistical operator  
$\rho=Z^{-1}\exp(-\beta{\cal K})$. 
To gain access to 
lower temperatures, we generalize the stochastic Gutzwiller Ansatz 
\cite{carusotto-NJPh-2003}, applying  
it to blocks of sites. 
Namely, assume for purposes of explanation, 
that the total grid of $M$ sites is divided into 
two blocks of $M/2$ sites each. 
Hamiltonian (\ref{discrH}) may be written as, 
$
\hat{\mathcal K}  = \Ham_1 + \Ham_2 + \Ham_{12} 
$, 
where 
%\begin{align} 
%\begin{aligned}
$
\Ham_{12}  = 
-\frac{\hbar^2}{2m\Delta x^2}
\hat b_{M/2}^\dagger\hat b_{M/2+1}
+ \text{h.c.} 
$, 
%\end{aligned}%
%% \nonumber % \eqlabel{} 
%\end{align}%
while $\Ham_1$ ($\Ham_2$) acts in the Hilbert space of the first (second) 
block.  
Applying the Trotter decomposition and the split approximation, 
we have, 
\begin{equation} 
e^{-\beta \hat{\mathcal K}}\approx 
\left[e^{-d\tau \Ham_{1}}e^{-d\tau \Ham_{2}}e^{-d\tau \Ham_{12}} 
\right]^A,\quad A\gg 1, \label{Trotter}
\end{equation} 
where $d\tau=\beta/A$ is the step in imaginary time. 
Let $\eta_{\tau}$ be $A$ independent 
complex Gaussian stochastic variables,  
one per each time step $\tau $, 
such that $\overline{\eta_{\tau}}=0$, $\overline{\eta^2_{\tau}}=0$, and
$\overline{\left|\eta_{\tau}
\right|^2} =
{\hbar^2 d\tau }/{2m\Delta x^2}  
$. ($\tau $ here is regarded as an index numbering steps rather than 
a continuous variable.) 
Then, 
%\begin{widetext}
% 
% 
\begin{align} 
\begin{aligned}
&e^{-d\tau\Ham_{12}} \approx 1 - d\tau\Ham_{12} = 
\\ 
&\ \ \overline{
\big(
1 - \eta_{\tau } \hat b_{M/2}^\dagger
+ \eta^*_{\tau } \hat b_{M/2}
\big) 
\big(
1 - \eta_{\tau } \hat b_{M/2+1}^\dagger
+ \eta^*_{\tau } \hat b_{M/2+1}
\big) 
}
, 
\end{aligned}%
\label{fact} 
\end{align}%
%\end{widetext}
and the evolution operator is ``stochastically 
factorised'': 
\begin{widetext}
\begin{align} 
\begin{aligned}
e^{-\beta \hat{\mathcal K}}\approx 
\overline{
\left\{
\prod_\tau
\left[e^{-d\tau \Ham_{1}} \big(
1 - \eta_{\tau } \hat b_{M/2}^\dagger
+ \eta^*_{\tau } \hat b_{M/2}
\big)
\right]\right\} 
\left\{
\prod_\tau
\left[e^{-d\tau \Ham_{2}} \big(
1 - \eta_{\tau } \hat b_{M/2+1}^\dagger
+ \eta^*_{\tau } \hat b_{M/2+1}
\big)
\right]\right\} 
}
. 
\end{aligned}%
% \nonumber % \eqlabel{} 
\end{align}%
\end{widetext}
(Recall that $\eta_{\tau }$ are independent 
for different $\tau $.)
Generalisation to arbitrary number of blocks is straightforward.

Introducing blocks we however face another problem: the  
blocks themselves should be sufficiently small to handle them  
numerically.  
Since the block 
Hilbert spaces are quite large even when restricting ourselves to 
small particle numbers,  
some truncation must be introduced.  
Choosing na\"ively  
the lowest eigenstates of the block Hamiltonians is dangerous  
because they do not necessarily correspond to the lowest  
eigenstates of the whole system.  
A fix to this problem is well known from the DMRG techniques  
\cite{Schollwoeck2004}.  
Following those ideas,   
we firstly build a thermal $\rho$-matrix of a block  
with one additional site at each end (environment). 
The most important states are then found by tracing out 
the environmental degrees of freedom,   
diagonalising the block $\rho$-matrix  
thus obtained, and finally  
taking the states corresponding to largest eigenvalues. 
 
Not unexpectedly, the convergence of the stochastic  
factorization method becomes worse when $\Delta x$ or the temperature 
are decreased. $\Delta x$ must be chosen small enough, for the 
discrete system to approximate well the continuous one;  
the number  
of particles per site must be small compared to unity  
\cite{paredes2004,Cazalilla2004}. 
Reducing $\Delta x$ leads to   
larger quasi-momentum cut-off and thus to larger noise in 
the simulation of the kinetic energy. 
With the block factorization method and $\Delta x < 1/n(0)$ we were 
able to reach a temperature of $k_\text{B} T  =\hbar \omega$, 
corresponding to $T \approx T_{\rm ph}$.  
Going to even lower temperature is mainly a question of a  
suitable Hilbert-space 
truncation.  
Block artifacts, i.e. effects of badly chosen block states,  
although much reduced by use of the environment, show 
up especially at low temperature. 
 
\begin{figure}[htb] 
\begin{center} 
\includegraphics[width=6.5cm]{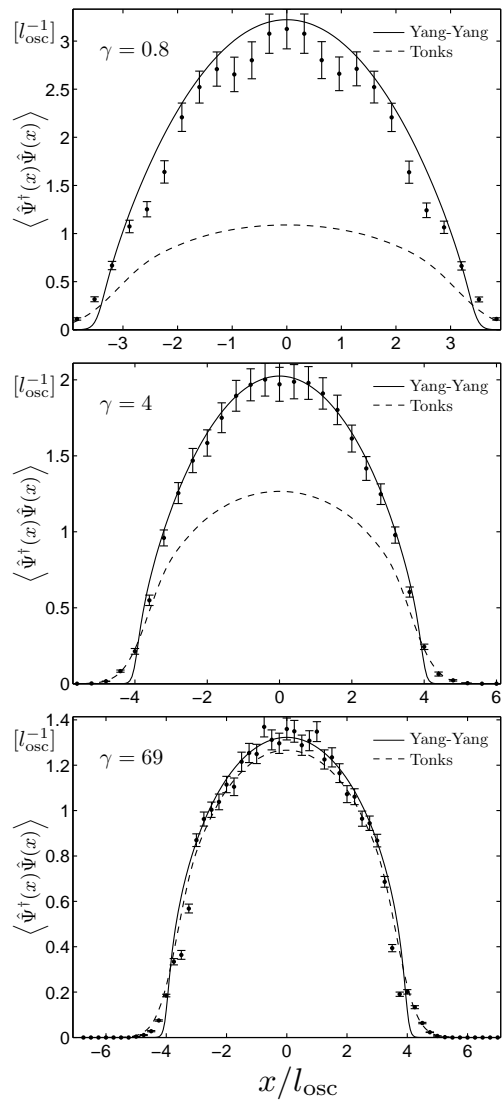} 
\caption{Particle density of the Bose gas in a trap at $k_\text{B}T=\hbar\omega$  
for different interaction strengths. Dots with error bars: stochastic simulation, solid line: prediction 
from Yang and Yang within the local density approximation, dashed line:  
Tonks (fermionization) limit.} 
\label{densitypics} 
\end{center} 
\end{figure}

\begin{figure}[htb] 
\begin{center} 
\includegraphics[width=6.5cm]{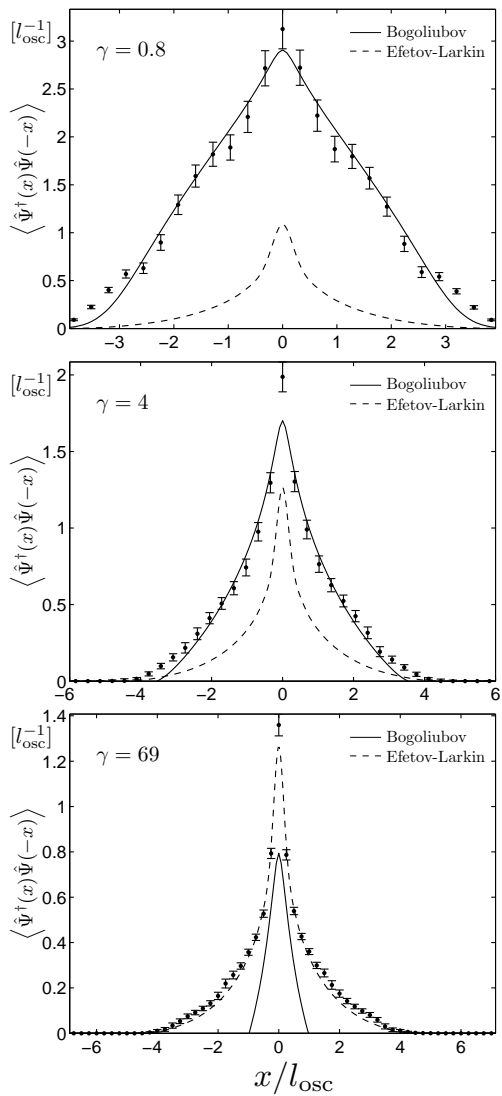} 
\caption{First-order correlations in the Bose gas for the same 
parameter regimes as in fig. \ref{densitypics}. Dots with error bars: stochastic simulation, solid line: Bogoliubov approximation, dashed line: Efetov-Larkin approximation.} 
\label{correlationpics} 
\end{center} 
\end{figure}

In figures \ref{densitypics} and \ref{correlationpics} numerical 
results for the density and the first order correlations in a 1D 
trap are shown for $k_\text{B} T = \hbar \omega$. The densities are compared 
to the predictions obtained from the Bethe-Ansatz solutions of  
Yang and Yang \cite{yang1968}  
using a local density approximation 
[i.e. replacing $\mu$ by $\mu(x)=\mu-V(x)$]. Also shown is  
the Tonks-gas limit, $\gamma=\infty$, at the 
given temperature, which is obtained using the mapping to a 
free Fermi gas. Apart from the case of $\gamma=0.8$  
where block artifacts are still present,  
there is a very good agreement with the 
prediction of the Yang-Yang theory in local density approximation.  
The latter becomes invalid close to the edges of the 
gas, and thus larger deviations of the 
numerical simulation from the Yang-Yang theory occur.

%%%%%%%%%%%%%%%%%%%%%%%%%%%%%%%%%%%%%%%%%%%%%%%%%%%%%%%%%%%%%%%%%%%%%%%%%%%% 
 
%%%%%%%%%%%%%%%%%%%%%%%%%%%%%%%%%%%%%%%%%%%%%%%%%%%%%%%%%%%%%%%%%%%%%%%%%%%%% 

%%%%%%%%%%%%%%%%%%%%%%%%%%%%%%%%%%%%%%%%%%%%%%%%%%%%%%%%%%%%%%%%%%%%%%%%%%%%% 
  
%%%%%%%%%%%%%%%%%%%%%%%%%%%%%%%%%%%%%%%%%%%%%%%%%%%%%%%%%%%%%%%%%%%%%%%%%%%% 

The Yang-Yang solution unfortunately giving no information 
about the correlations in the system, we compare the corresponding numerical results 
to different predictions valid either in the weak or strong interaction limits. 
The weak-interaction limit is described by a Bogoliubov approximation. 
Since the temperature is rather low we here have not taken into 
account thermal depletion of the quasi-condensate.  
% The latter could be 
% done, however, within a Popov approach \cite{Popov}. 
In the opposite limit, $\gamma\to\infty$, 
correlations can be calculated by mapping 
impenetrable bosons $\hat b_i, \hat b_i^\dagger$ to  
fermions $\hat c_i, \hat c_i^\dagger$ via   
% 
% 
%\be 
$\hat b_i =\prod_{j<i} (1-2 \hat c_j^\dagger \hat c_j)\, \hat c_i$, 
%\ee 
% 
% 
which leads to the expression for first-order correlations 
found by Efetov and Larkin \cite{efetov1975} 
\be 
\langle \hat b_i^\dagger \hat b_j\rangle = {\rm Det} 
\Bigl(\mathbf{g}^{ij}\Bigr),\qquad i< j. 
\ee 
$\mathbf{g}^{ij}$ is a $(j-i)\times(j-i)$ matrix with elements 
% 
% 
%\be 
$(g^{ij})_{n,m}=\langle \hat c_n^\dagger \hat c_{m+1}\rangle -\delta_{nm+1}/2$, 
%\ee 
% 
% 
with $n$ and $m$ running from $i$ to $j-1$.  
Fig. \ref{correlationpics} compares the simulated correlations with the 
Bogoliubov and Efetov-Larkin predictions.  
The expected transition from the Bogoliubov to the 
Efetov-Larkin behavior in the Tonks limit with increasing $\gamma$ 
is clearly seen.  
 
\begin{figure}[ht] 
\begin{center} 
\includegraphics[width=6cm]{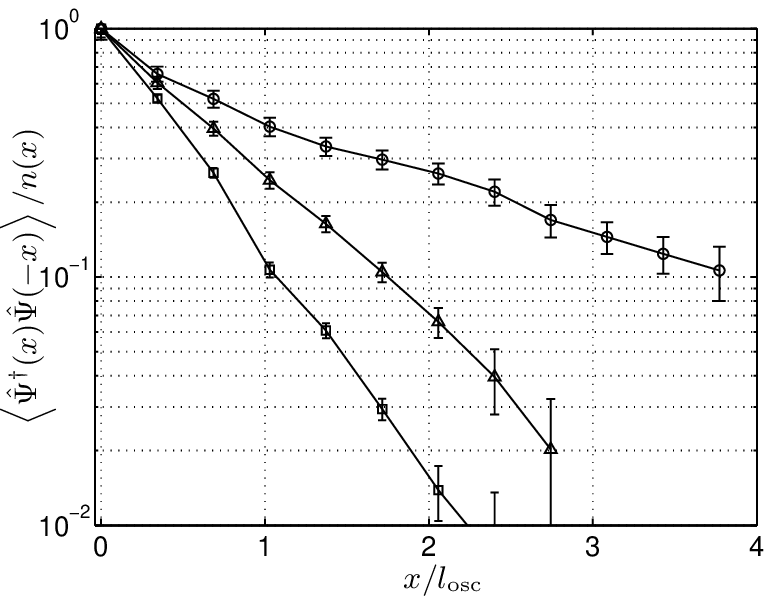} 
\caption{First-order correlations for $\gamma=4$ at different temperatures  
shown on a logarithmic scale:  
$k_\text{B} T/\hbar\omega = 3 $ (squares), $k_\text{B} T/\hbar\omega = 2$ (triangles),  
and $k_\text{B} T/\hbar\omega = 1$ (circles).  
%One recognizes exponential decay with increasing correlation 
%length for decreasing temperature. 
At $k_\text{B}T= \hbar\omega$, a deviation from the  
exponential behavior becomes discernible.} 
\label{log+loglog} 
\end{center} 
\end{figure}

To see the asymptotic behavior of the phase correlations more clearly,  
in fig. \ref{log+loglog} we plot  
the first-order correlations  
normalized to the density $\langle \hat \Psi(x)^\dagger\hat\Psi(-x)\rangle / n(x)$ 
for $\gamma=4$  on a logarithmic scale.  
We show numerical results for three different temperatures,  
$k_\text{B} T/\hbar\omega=1, 2, 3$. 
For the lowest temperature  we were able to 
reach in our simulation, deviations 
from the pure exponential decay characteristic of higher temperatures  
can already  
be seen for intermediate distances.  
This is consistent with predictions of the Luttinger-liquid  
model \cite{Cazalilla-JPhysB-2004}, namely, that the asymptotic  
behaviour of the correlations  
changes from exponential to a power law if the thermal energy $k_\text{B} T$ becomes much smaller  
than the trap energy $\hbar\omega$.  
(The spatial resolution of our simulations 
is insufficient to see the short-distance behavior of the correlations 
which is not described by the LL model.)  
In order to go to even lower temperatures,
we have also employed a finite-temperature variant of the DMRG method 
\cite{Plimak-PRA-2004}. This method works well for the complementary 
regime of very low temperatures $k_BT/\hbar\omega < 0.1$. The results 
show a smooth continuation of the behavior found in the present paper 
and will be discussed elsewhere.
 
%%%%%%%%%%%%%%%%%%%%%%%%%%%%%%%%%%%%%%%%%%%%%%%%%%%%%%%%%%%%%%%%%%%%%%%%%%%% 
 
%%%%%%%%%%%%%%%%%%%%%%%%%%%%%%%%%%%%%%%%%%%%%%%%%%%%%%%%%%%%%%%%%%%%%%%%%%%%% 
 
In summary 
we have presented a numerical method for the stochastic simulation 
of thermal properties of one dimensional Bose gases covering the  
whole range from 
the Bogoliubov regime of weak interactions to the Tonks-Girardeau 
regime of strong interactions. The method is based on a generalization  
of the stochastic Gutzwiller Ansatz to blocks of sites in the discretized 
Hamiltonian. Combining the stochastic block factorization scheme with  
a suitable selection of the most important states of the block,  
we were able to reach the transition regime between 
thermal and quantum dominated correlations. 
%%%%%%%%%%%%%%%%%%%%%%%%%%%%%%%%%%%%%%%%%%%%%%%%%%%%%%%%%%%%%%
%
%

The financial support of the Deutsche Forschungsgemeinschaft  
through the SPP 1116 ``Ultrakalte Quantengase'' is gratefully acknowledged.

\bibliography{biblio} 

\begin{thebibliography}{27}
\expandafter\ifx\csname natexlab\endcsname\relax\def\natexlab#1{#1}\fi
\expandafter\ifx\csname bibnamefont\endcsname\relax
  \def\bibnamefont#1{#1}\fi
\expandafter\ifx\csname bibfnamefont\endcsname\relax
  \def\bibfnamefont#1{#1}\fi
\expandafter\ifx\csname citenamefont\endcsname\relax
  \def\citenamefont#1{#1}\fi
\expandafter\ifx\csname url\endcsname\relax
  \def\url#1{\texttt{#1}}\fi
\expandafter\ifx\csname urlprefix\endcsname\relax\def\urlprefix{URL }\fi
\providecommand{\bibinfo}[2]{#2}
\providecommand{\eprint}[2][]{\url{#2}}

\bibitem[{\citenamefont{Lieb and Liniger}(1963)}]{lieb1963}
\bibinfo{author}{\bibfnamefont{E.~H.} \bibnamefont{Lieb}} \bibnamefont{and}
  \bibinfo{author}{\bibfnamefont{W.}~\bibnamefont{Liniger}},
  \bibinfo{journal}{Phys. Rev.} \textbf{\bibinfo{volume}{130}},
  \bibinfo{pages}{1605} (\bibinfo{year}{1963}).

\bibitem[{\citenamefont{Yang and Yang}(1969)}]{yang1968}
\bibinfo{author}{\bibfnamefont{C.~N.} \bibnamefont{Yang}} \bibnamefont{and}
  \bibinfo{author}{\bibfnamefont{C.~P.} \bibnamefont{Yang}},
  \bibinfo{journal}{J. Math. Phys.} \textbf{\bibinfo{volume}{10}},
  \bibinfo{pages}{1115} (\bibinfo{year}{1969}).

\bibitem[{\citenamefont{Korepin et~al.}(1993)\citenamefont{Korepin, Bogoliubov,
  and Izergin}}]{bogoliubov-book}
\bibinfo{author}{\bibfnamefont{V.~E.} \bibnamefont{Korepin}},
  \bibinfo{author}{\bibfnamefont{N.~M.} \bibnamefont{Bogoliubov}},
  \bibnamefont{and} \bibinfo{author}{\bibfnamefont{A.~G.}
  \bibnamefont{Izergin}}, \emph{\bibinfo{title}{Quantum Inverse Scattering
  Method and Correlation Functions}} (\bibinfo{publisher}{Cambridge University
  Press}, \bibinfo{year}{1993}).

\bibitem[{\citenamefont{Paredes et~al.}(2004)\citenamefont{Paredes, Widera,
  Murg, Mandel, F\"olling, Cirac, Shlyapnikov, H\"ansch, and
  Bloch}}]{paredes2004}
\bibinfo{author}{\bibfnamefont{B.}~\bibnamefont{Paredes}},
  \bibinfo{author}{\bibfnamefont{A.}~\bibnamefont{Widera}},
  \bibinfo{author}{\bibfnamefont{V.}~\bibnamefont{Murg}},
  \bibinfo{author}{\bibfnamefont{O.}~\bibnamefont{Mandel}},
  \bibinfo{author}{\bibfnamefont{S.}~\bibnamefont{F\"olling}},
  \bibinfo{author}{\bibfnamefont{I.}~\bibnamefont{Cirac}},
  \bibinfo{author}{\bibfnamefont{G.~V.} \bibnamefont{Shlyapnikov}},
  \bibinfo{author}{\bibfnamefont{T.~W.} \bibnamefont{H\"ansch}},
  \bibnamefont{and} \bibinfo{author}{\bibfnamefont{I.}~\bibnamefont{Bloch}},
  \bibinfo{journal}{Nature} \textbf{\bibinfo{volume}{429}},
  \bibinfo{pages}{277} (\bibinfo{year}{2004}).

\bibitem[{\citenamefont{Kinoshita et~al.}(2004)\citenamefont{Kinoshita, Wenger,
  and Weiss}}]{Kinoshita2004}
\bibinfo{author}{\bibfnamefont{T.}~\bibnamefont{Kinoshita}},
  \bibinfo{author}{\bibfnamefont{T.}~\bibnamefont{Wenger}}, \bibnamefont{and}
  \bibinfo{author}{\bibfnamefont{D.~S.} \bibnamefont{Weiss}},
  \bibinfo{journal}{Science} \textbf{\bibinfo{volume}{305}},
  \bibinfo{pages}{1125} (\bibinfo{year}{2004}).

\bibitem[{\citenamefont{Olshanii}(1998)}]{olshanii1998}
\bibinfo{author}{\bibfnamefont{M.}~\bibnamefont{Olshanii}},
  \bibinfo{journal}{Phys. Rev. Lett.} \textbf{\bibinfo{volume}{81}},
  \bibinfo{pages}{938} (\bibinfo{year}{1998}).

\bibitem[{\citenamefont{Hohenberg}(1967)}]{Hohenberg1967}
\bibinfo{author}{\bibfnamefont{P.~C.} \bibnamefont{Hohenberg}},
  \bibinfo{journal}{Phys. Rev.} \textbf{\bibinfo{volume}{158}},
  \bibinfo{pages}{383} (\bibinfo{year}{1967}).

\bibitem[{\citenamefont{Efetov and Larkin}(1975)}]{efetov1975}
\bibinfo{author}{\bibfnamefont{K.~B.} \bibnamefont{Efetov}} \bibnamefont{and}
  \bibinfo{author}{\bibfnamefont{A.~I.} \bibnamefont{Larkin}},
  \bibinfo{journal}{Sov. Phys.-JETP} \textbf{\bibinfo{volume}{42}},
  \bibinfo{pages}{390} (\bibinfo{year}{1975}).

\bibitem[{\citenamefont{Haldane}(1981)}]{Haldane1981}
\bibinfo{author}{\bibfnamefont{F.~D.~M.} \bibnamefont{Haldane}},
  \bibinfo{journal}{Phys. Rev. Lett.} \textbf{\bibinfo{volume}{47}},
  \bibinfo{pages}{1840} (\bibinfo{year}{1981}).

\bibitem[{\citenamefont{Tonks}(1936)}]{Tonks1936}
\bibinfo{author}{\bibfnamefont{L.}~\bibnamefont{Tonks}},
  \bibinfo{journal}{Phys. Rev.} \textbf{\bibinfo{volume}{50}},
  \bibinfo{pages}{955} (\bibinfo{year}{1936}).

\bibitem[{\citenamefont{Girardeau}(1960)}]{Girardeau1960}
\bibinfo{author}{\bibfnamefont{M.}~\bibnamefont{Girardeau}},
  \bibinfo{journal}{J. Math. Phys.} \textbf{\bibinfo{volume}{1}},
  \bibinfo{pages}{516} (\bibinfo{year}{1960}).

\bibitem[{\citenamefont{Kheruntsyan et~al.}(2003)\citenamefont{Kheruntsyan,
  Gangardt, Drummond, and Shlyapnikov}}]{Kheruntsyan2003}
\bibinfo{author}{\bibfnamefont{K.~V.} \bibnamefont{Kheruntsyan}},
  \bibinfo{author}{\bibfnamefont{D.~M.} \bibnamefont{Gangardt}},
  \bibinfo{author}{\bibfnamefont{P.~D.} \bibnamefont{Drummond}},
  \bibnamefont{and} \bibinfo{author}{\bibfnamefont{G.~V.}
  \bibnamefont{Shlyapnikov}}, \bibinfo{journal}{Phys. Rev. Lett.}
  \textbf{\bibinfo{volume}{91}}, \bibinfo{pages}{040403}
  (\bibinfo{year}{2003}).

\bibitem[{\citenamefont{Petrov et~al.}(2000)\citenamefont{Petrov, Shlyapnikov,
  and Walraven}}]{petrov2000}
\bibinfo{author}{\bibfnamefont{D.~S.} \bibnamefont{Petrov}},
  \bibinfo{author}{\bibfnamefont{G.~V.} \bibnamefont{Shlyapnikov}},
  \bibnamefont{and} \bibinfo{author}{\bibfnamefont{J.~T.~M.}
  \bibnamefont{Walraven}}, \bibinfo{journal}{Phys. Rev. Lett.}
  \textbf{\bibinfo{volume}{85}}, \bibinfo{pages}{3745} (\bibinfo{year}{2000}).

\bibitem[{\citenamefont{Luxat and Griffin}(2003)}]{luxat2003}
\bibinfo{author}{\bibfnamefont{D.~L.} \bibnamefont{Luxat}} \bibnamefont{and}
  \bibinfo{author}{\bibfnamefont{A.}~\bibnamefont{Griffin}},
  \bibinfo{journal}{Phys. Rev. A} \textbf{\bibinfo{volume}{67}},
  \bibinfo{pages}{043603} (\bibinfo{year}{2003}).

\bibitem[{\citenamefont{Olshanii and Dunjko}(2003)}]{olshanii2003}
\bibinfo{author}{\bibfnamefont{M.}~\bibnamefont{Olshanii}} \bibnamefont{and}
  \bibinfo{author}{\bibfnamefont{V.}~\bibnamefont{Dunjko}},
  \bibinfo{journal}{Phys. Rev. Lett.} \textbf{\bibinfo{volume}{91}},
  \bibinfo{pages}{090401} (\bibinfo{year}{2003}).

\bibitem[{\citenamefont{Cazalilla}(2004)}]{Cazalilla-JPhysB-2004}
\bibinfo{author}{\bibfnamefont{M.~A.} \bibnamefont{Cazalilla}},
  \bibinfo{journal}{J. Phys. B} \textbf{\bibinfo{volume}{37}},
  \bibinfo{pages}{S1} (\bibinfo{year}{2004}).

\bibitem[{\citenamefont{Monien et~al.}(1998)\citenamefont{Monien, Linn, and
  Elstner}}]{Monien1998}
\bibinfo{author}{\bibfnamefont{H.}~\bibnamefont{Monien}},
  \bibinfo{author}{\bibfnamefont{M.}~\bibnamefont{Linn}}, \bibnamefont{and}
  \bibinfo{author}{\bibfnamefont{N.}~\bibnamefont{Elstner}},
  \bibinfo{journal}{Phys. Rev. A} \textbf{\bibinfo{volume}{58}},
  \bibinfo{pages}{R3395} (\bibinfo{year}{1998}).

\bibitem[{\citenamefont{Gangardt and Shlyapnikov}(2003)}]{Gangardt2003}
\bibinfo{author}{\bibfnamefont{D.~M.} \bibnamefont{Gangardt}} \bibnamefont{and}
  \bibinfo{author}{\bibfnamefont{G.~V.} \bibnamefont{Shlyapnikov}},
  \bibinfo{journal}{Phys. Rev. Lett.} \textbf{\bibinfo{volume}{90}},
  \bibinfo{pages}{010401} (\bibinfo{year}{2003}).

\bibitem[{\citenamefont{Plimak et~al.}(2003)\citenamefont{Plimak, Schmidt, and
  Fleischhauer}}]{Plimak-DPG-2003}
\bibinfo{author}{\bibfnamefont{L.}~\bibnamefont{Plimak}},
  \bibinfo{author}{\bibfnamefont{B.}~\bibnamefont{Schmidt}}, \bibnamefont{and}
  \bibinfo{author}{\bibfnamefont{M.}~\bibnamefont{Fleischhauer}}
  (\bibinfo{year}{2003}), \bibinfo{note}{spring meeting DPG, Q 27.15}.

\bibitem[{\citenamefont{Carusotto and
  Castin}(2003{\natexlab{a}})}]{carusotto-NJPh-2003}
\bibinfo{author}{\bibfnamefont{I.}~\bibnamefont{Carusotto}} \bibnamefont{and}
  \bibinfo{author}{\bibfnamefont{Y.}~\bibnamefont{Castin}},
  \bibinfo{journal}{New Journal of Phys.} \textbf{\bibinfo{volume}{5}},
  \bibinfo{pages}{91.1} (\bibinfo{year}{2003}{\natexlab{a}}).

\bibitem[{\citenamefont{Schollwoeck}(2005)}]{Schollwoeck2004}
\bibinfo{author}{\bibfnamefont{U.}~\bibnamefont{Schollwoeck}},
  \bibinfo{journal}{Rev. Mod. Phys.}  (\bibinfo{year}{2005}),
  \bibinfo{note}{cond-mat/0409292}.

\bibitem[{\citenamefont{Carusotto and
  Castin}(2003{\natexlab{b}})}]{Carusotto-PRL-2003}
\bibinfo{author}{\bibfnamefont{I.}~\bibnamefont{Carusotto}} \bibnamefont{and}
  \bibinfo{author}{\bibfnamefont{Y.}~\bibnamefont{Castin}},
  \bibinfo{journal}{Phys. Rev. Lett.} \textbf{\bibinfo{volume}{90}},
  \bibinfo{pages}{030401} (\bibinfo{year}{2003}{\natexlab{b}}).

\bibitem[{\citenamefont{Drummond and Gardiner}(1980)}]{drummond1980}
\bibinfo{author}{\bibfnamefont{P.~D.} \bibnamefont{Drummond}} \bibnamefont{and}
  \bibinfo{author}{\bibfnamefont{C.~W.} \bibnamefont{Gardiner}},
  \bibinfo{journal}{J. Phys. A: Math. Gen.} \textbf{\bibinfo{volume}{13}},
  \bibinfo{pages}{2353} (\bibinfo{year}{1980}).

\bibitem[{\citenamefont{Plimak et~al.}(1998)\citenamefont{Plimak, Fleischhauer,
  and Walls}}]{plimak1998}
\bibinfo{author}{\bibfnamefont{L.~I.} \bibnamefont{Plimak}},
  \bibinfo{author}{\bibfnamefont{M.}~\bibnamefont{Fleischhauer}},
  \bibnamefont{and} \bibinfo{author}{\bibfnamefont{D.~F.} \bibnamefont{Walls}},
  \bibinfo{journal}{Europhys. Lett.} \textbf{\bibinfo{volume}{43}},
  \bibinfo{pages}{641} (\bibinfo{year}{1998}).

\bibitem[{\citenamefont{Drummond et~al.}(2004)\citenamefont{Drummond, Deuar,
  and Kheruntsyan}}]{Drummond2004}
\bibinfo{author}{\bibfnamefont{P.~D.} \bibnamefont{Drummond}},
  \bibinfo{author}{\bibfnamefont{P.}~\bibnamefont{Deuar}}, \bibnamefont{and}
  \bibinfo{author}{\bibfnamefont{K.~V.} \bibnamefont{Kheruntsyan}},
  \bibinfo{journal}{Phys. Rev. Lett.} \textbf{\bibinfo{volume}{92}},
  \bibinfo{pages}{040405} (\bibinfo{year}{2004}).

\bibitem[{\citenamefont{Cazalilla}()}]{Cazalilla2004}
\bibinfo{author}{\bibfnamefont{M.~A.} \bibnamefont{Cazalilla}},
  \bibinfo{note}{cond-mat/0406526}.

\bibitem[{\citenamefont{Plimak et~al.}(2004)\citenamefont{Plimak, Olsen, and
  Fleischhauer}}]{Plimak-PRA-2004}
\bibinfo{author}{\bibfnamefont{L.~I.} \bibnamefont{Plimak}},
  \bibinfo{author}{\bibfnamefont{M.~K.} \bibnamefont{Olsen}}, \bibnamefont{and}
  \bibinfo{author}{\bibfnamefont{M.}~\bibnamefont{Fleischhauer}},
  \bibinfo{journal}{Phys. Rev. A} \textbf{\bibinfo{volume}{70}},
  \bibinfo{pages}{013611} (\bibinfo{year}{2004}).

\end{thebibliography}
 
\end{document}